\documentclass[12pt,preprint]{aastex}

\shorttitle{DLA-LBG spatial cross-correlation at $z\sim3$}
\shortauthors{Cooke et al.}
 
\begin{document}

\title{Measurement of the Spatial Cross-Correlation Function of 
  Damped Lyman $\alpha$ Systems and Lyman Break Galaxies}

\author{Jeff Cooke\altaffilmark{1} \altaffilmark{2}
\& Arthur M. Wolfe\altaffilmark{1}}
\affil{Department of Physics and Center for Astrophysics and
  Space Sciences; University of California, San Diego, mc-0424, La
  Jolla, CA 92093-0424}
\email{cooke@physics.ucsd.edu}
\email{awolfe@ucsd.edu}
 
\author{Eric Gawiser\altaffilmark{1}\altaffilmark{3}}
\affil{Department of Astronomy; Yale University, P.O. Box 208101, New
  Haven, CT 06520-8101}
\email{gawiser@astro.yale.edu}

\and

\author{Jason X. Prochaska\altaffilmark{1}}
\affil{UCO-Lick Observatory; University of California, Santa Cruz,
  Santa Cruz, CA 95064}
\email{xavier@ucolick.edu}

\altaffiltext{1}{Visiting Astronomer, W. M. Keck Telescope. The Keck
  Observatory is a joint facility of the University of California, the
  California Institute of Technology, and NASA and was made possible by 
  the generous financial support of the W. M. Keck Foundation.}
\altaffiltext{2}{Center for Cosmology, University of California,
  Irvine, CA 92697; cooke@uci.edu}
\altaffiltext{3}{NSF Astronomy \& Astrophysics Postdoctoral Fellow}

\begin{abstract}
We present the first spectroscopic measurement of the spatial
cross-correlation function between damped Lyman $\alpha$ systems
(DLAs) and Lyman break galaxies (LBGs).  We obtained deep $u'$BVRI
images of nine QSO fields with 11 known $z\sim3$ DLAs and 
spectroscopically confirmed 211 R $<25.5$ photometrically selected
$z>2$ LBGs.  We find strong evidence for an overdensity of LBGs near
DLAs versus random, the results of which are similar to that of LBGs
near other LBGs.  A maximum likelihood cross-correlation analysis
found the best fit correlation length value of
$r_0=2.9^{+1.4}_{-1.5}h^{-1}$Mpc using a fixed value of $\gamma=1.6$.  
The implications of the DLA-LBG clustering amplitude on the average
dark matter halo mass of DLAs are discussed.
\end{abstract}


\section{Introduction}

Over the last two decades, quasar (QSO) absorption line systems have
provided tremendous insight into the nature of proto-galaxies at high
redshift.  Of particular interest are the damped Lyman $\alpha$
systems (DLAs) defined to have N(\textsc{Hi}) $\ge 2\times10^{20}$
atoms cm$^{-2}$ \citep{w86,w05}.  Systems with such large column
densities can provide self-shielding from the ambient ionizing
radiation at high redshift and protect large reservoirs of neutral
gas.  Such systems are prime sites for star formation.  Moderate and 
high-resolution databases \citep{p01,p03} and numerous spectroscopic 
studies of DLAs have given us a detailed view of the chemical
abundances and gas kinematics of proto-galaxies, yet, except for two 
confirmed $z>2$ detections of DLA emission \citep{m02,m04},
information regarding their mass, luminosity, and morphology has
remained elusive.

One approach to measure the average mass of a galaxy population
at high redshift is under the implicit assumption that galaxies are a
result of the gravitational instabilities of primordial density
fluctuations.  CDM hierarchical models predict that the most massive
galaxies at high redshift preferentially form clustered together near
the density peaks of regions with an underlying mass overdensity,
whereas low-mass galaxies form more uniformly throughout space 
\citep{k84,b86}.  The factor in which the underlying dark matter is 
enhanced in the regions where galaxies cluster as compared to that 
implied by the galaxies themselves is referred to as bias.  In this
context, the spatial distribution of a population of galaxies provides
a means to infer their typical dark matter halo mass.  This method has
been used to infer the average mass of Lyman break galaxies (LBGs) at
$z\sim3$ \citep{s98,a98} and agrees with the mass estimates from
nebular line-width measurements \citep{pet01} and those implied by
star formation model fits \citep{aes01}.

It is difficult to measure the mass of DLAs by their spatial
clustering because of the sparse distribution of bright QSO sightlines
and since only approximately one quarter of all $z>3$ QSOs exhibit DLA 
absorption. However, the mass of DLAs can be inferred by their 
cross-correlation with another known population \citep{g01}.  Since
the LBG auto-correlation function and LBG galaxy bias at $z\sim3$ has
been established \citep{a03}, it is natural to use these galaxies as
tracers of the underlying mass distribution to cross-correlate with 
$z\sim3$ DLAs.  With this in mind, A03 used the spectroscopic sample
of \citet{s03} to test the spatial distribution of LBGs and DLAs.  A
count of the number of LBGs in three-dimensional cells centered on the
four DLAs in their sample showed no significant overdensity.  In
contrast, \citet{b04} used photometric redshifts to measure the
clustering of LBGs near two DLAs and one sub-DLA in wide-field images.  
From Monte Carlo simulations, they found a non-zero DLA-LBG clustering 
amplitude to greater than $2\sigma$ and angular analysis on scales out
to $\sim20h^{-1}$Mpc estimated the DLA-LBG cross-correlation to be
equal to, or greater than, the LBG auto-correlation.  Both analyses
had limited statistics and could neither confirm, nor rule out, a
significant overdensity of LBGs near DLAs.

In this Letter, we highlight the results of a spectroscopic survey for
LBGs associated with 11 DLAs at $z\sim3$.  We introduce strong
evidence for an overdensity of LBGs near DLAs and present the first 
detection of the three-dimensional DLA-LBG cross-correlation function.  
A more complete discussion of the DLA-LBG cross-correlation analysis
is presented in \citet{c05b} along with our independent measurement of
the $z\sim3$ LBG auto-correlation function.  In this Letter, we adopt 
$\Omega_{M}=0.3, \Omega_{\Lambda}=0.7$ cosmology.

\section{Observations} 

We acquired deep $u'$BVRI images from 2000 April through 2003 November
of nine QSO fields with 11 known DLAs $(2.78<z<3.32)$ using the Carnegie 
Observatories Spectrograph and Multi-Object Imaging Camera \citep{k98}
on the $200''$ Hale telescope at Palomar and the Low Resolution Imager
and Spectrometer \citep {o95} on the Keck I telescope.  The data were
reduced in a standard manner.  We developed a $u'$BVRI photometric
selection technique for LBGs at $z\sim3$ that proved comparable to
previous techniques in both efficiency and resulting redshift
distribution.  Over the 465 arcmin$^2$ surveyed, we found 796 R
$<25.5$ objects that met our color criteria.  Follow-up multi-object 
spectroscopy of 529 LBG candidates using LRIS yielded 339
redshifts. We identified 211 LBGs with $z>2$ and used these in the 
cross-correlation analysis.  Details of data acquisition, reduction,
and analysis can be found in \citet{c05a}.

\section{Clustering Analysis}

\subsection{Evidence of an LBG overdensity near DLAs}

As a coarse measure of the distribution of LBGs near DLAs, we divided
our survey volume into cells with dimensions of the field area of LRIS
at $z\sim3$ ($\sim7 \times 10 h^{-1}$Mpc) and $\Delta z=0.025 
~(\sim17h^{-1}$Mpc).  The choice in cell size follows that of 
\citet{a98} and includes the majority of the objects associated with a 
central object having a galaxy bias less than or equal to the LBG bias
at $z\sim3$.  The extended length in the redshift direction is
intended to account for the $\sim1-2h^{-1}$Mpc error in the systemic 
redshift measurement inherent to LBGs. 

This simple counts-in-cells analysis found an average of 1.27 objects
residing in cells centered on each of the 11 DLAs where an average of
0.85 objects should have been found randomly.  Random values were
determined for objects in identical cells at the redshifts of the
DLAs pulled from normalized random catalogs that mimicked the
constraints of the data and were corrected by the photometric
selection function [see \citet{c05a,c05b}].  This observed overdensity
can be compared to an average of 1.16 objects found in cells of
identical size centered on LBGs in our survey having similar redshifts 
to the DLAs but located in other fields.  Interestingly, two of the 14 
objects associated with the DLAs are QSOs.  Since QSOs are believed to 
form in massive dark matter halos that seed supermassive black holes, 
this suggests that the corresponding DLAs reside in overdense regions.  

\subsection{DLA-LBG cross-correlation function}

We measured the DLA-LBG cross-correlation function $\xi_{DLA-LBG}$
using the usual approach of comparing galaxy pair separations in the
data to galaxy pair separations in the random galaxy catalogs.  We
used the estimator of \citet{ls93} to measure the excess probability
over random of finding an LBG at a distance $r$ from a DLA  
\begin {equation}\label{DLAest}
\xi_{DLA-LBG}~(r) =
\frac{D_{DLA}D_{LBG}-D_{DLA}R_{LBG}-R_{DLA}D_{LBG}+R_{DLA}R_{LBG}}
{R_{DLA}R_{LBG}}
\end {equation}
\noindent where $D_{DLA}D_{LBG}$ is the catalog of data-data pair
separations, $D_{DLA}R_{LBG}$ and $R_{DLA}D_{LBG}$ are the data-random
and random-data pair separation cross-reference catalogs, and
$R_{DLA}R_{LBG}$ is the catalog of random-random pair separations.
This estimator is well-suited for small galaxy samples and has a
nearly Poisson variance.  The random catalogs were constructed to be
many times larger than the data catalog to reduce shot noise and were
then normalized to the data.  The mean LBG density was determined from
the data in all 11 fields.  We determined $\xi(r)$ by counting the
number of pairs in each catalog over a series of log or linear
intervals (i.e. bins).  In addition, we made the assumption that
$\xi(r)$ follows a power law of the form 
\begin {equation}\label{powerlaw}
\xi(r) = \left(\frac{r}{r_0}\right)^{-\gamma}.
\end {equation}

\subsection{Conventional binning}

We initially measured the cross-correlation function by duplicating
the cylindrical binning technique described in A03, Appendix C.  This
technique was adopted to help minimize the effect that LBG redshift 
uncertainties have on the clustering signal as compared to traditional 
radial bins.  In addition, this approach permitted a direct comparison
of our results to the published LBG auto-correlation values of A03
using the available online dataset\footnote{
http://vizier.cfa.harvard.edu/viz-bin/VizieR?-source=J/ApJ/592/728/}
of \citet{s03} since both surveys were executed in a similar manner and
used the same instruments and configurations.  

\begin{figure}
\begin{center}
%\scalebox{0.6}[0.6]{\rotatebox{90}{\includegraphics{f1.ps}}}
\scalebox{0.6}[0.6]{\rotatebox{90}{\includegraphics{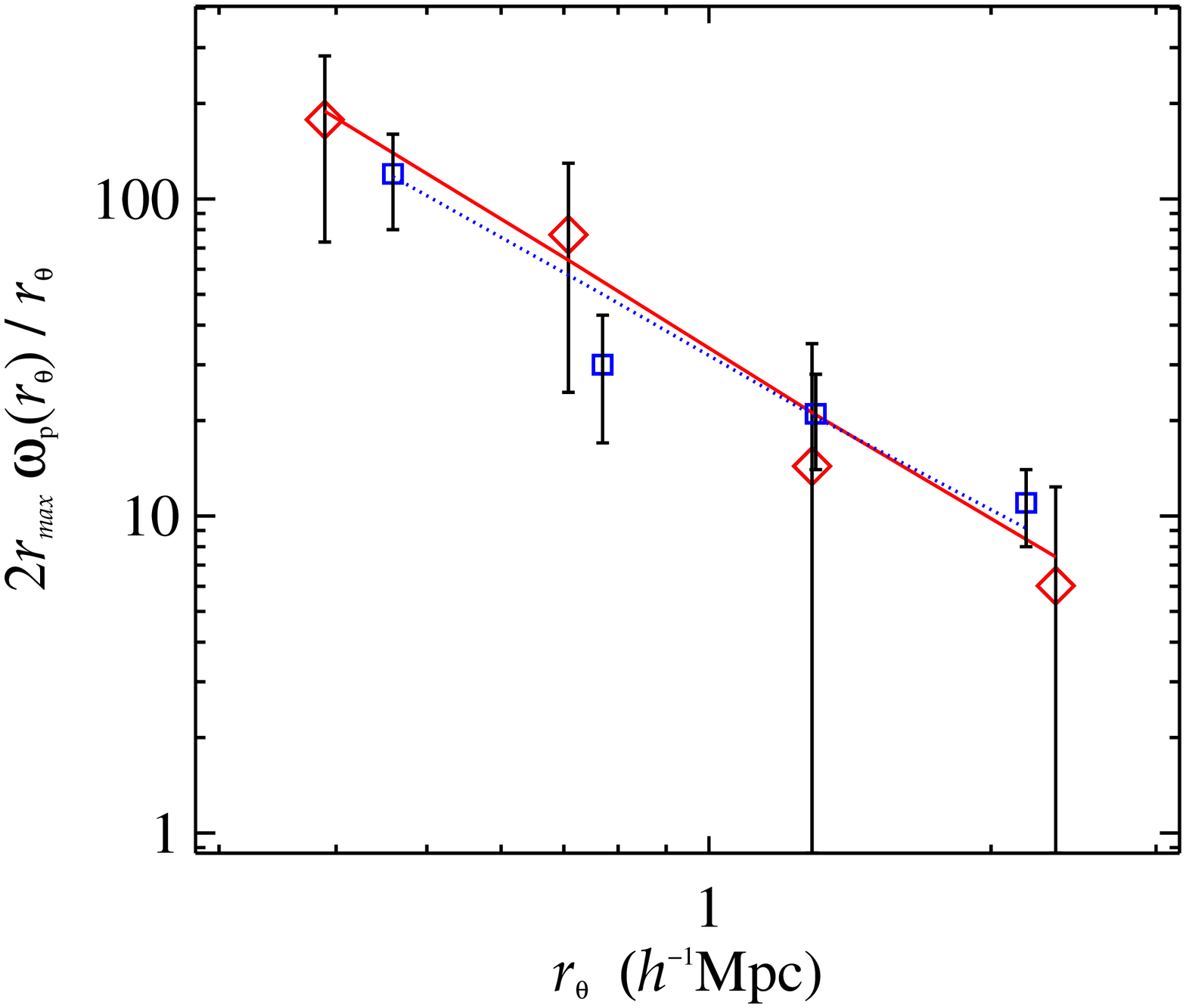}}}
\caption
{Measurement of the DLA-LBG cross-correlation following the binning
  method of Adelberger et al. (2003) and plotted in a consistent
  manner.  The cross-correlation values are indicated by red diamonds
  and we find a best fit of $r_0=3.3\pm1.3h^{-1}$Mpc$,
  ~\gamma=1.7\pm0.4$ indicated by the solid red line.  The errors
  shown are near Poisson and the reported errors are where 68\% of the
  best fit values lie
  from a Monte Carlo analysis of the functional fit.  For a fixed
  value of $\gamma=1.6$, we find a best fit correlation length of
  $r_0=3.5\pm1.0h^{-1}$Mpc.  The LBG auto-correlation (blue squares)
  of Adelberger et al. (2003) are overlaid over a similar scale with
  the published fit of $r_0=3.96\pm0.29h^{-1}$Mpc,
  $~\gamma=1.55\pm0.15$ (blue dotted line).  The DLA-LBG
  cross-correlation values are consistent with the angular wide-field
  analysis of \citet{b04} significant on scales of
  $\sim5-10h^{-1}$Mpc.}
\label{fig:omegap}
\end{center}
\end{figure}

In this treatment, the expected projected angular overdensity is
defined to be
\begin{equation}\label{omegap}
\omega_p(r_\theta)\equiv \frac{r_0^\gamma r_\theta^{1-\gamma}}{2r_z} B
\left(\frac{1}{2},\frac{\gamma - 1}{2} \right) I_x
\left(\frac{1}{2},\frac{\gamma - 1}{2} \right) 
\end{equation}
where $r_z$ is the greater of 1000 km sec$^{-1}(1+z)/H(z)$ and $7
r_\theta$, and $B$ and $I_x$ are the beta and incomplete beta
functions with $x\equiv r_z^2 (r_z^2 + r_\theta^2)^{-1}$ \citep{press92}. 
Applying this method to the DLA-LBG cross-correlation, we
found best fit parameter values and 1 sigma uncertainties of 
$r_0=3.3\pm1.3h^{-1}$Mpc,$~\gamma=1.7\pm0.4$.  Figure~\ref{fig:omegap}
presents and compares these results with the LBG auto-correlation
results of A03 and is plotted in a consistent manner where
$r_{max}=r_z$ as described above.  The errors on the
cross-correlation values shown in the figure are those determined
using the formulation in \citet{ls93} and the reported errors on the 
functional fit were determined by duplicating the Monte Carlo error
analysis as described in A03. The latter error analysis may 
underestimate the true error by a factor of $\sim1-2$ \citep{a05}.

Although the uncertainties are large, it is immediately apparent from 
Figure~\ref{fig:omegap} that the form and central values of the two 
correlation functions are similar.  In addition, we computed a 
cross-correlation length of $r_0=3.5\pm1.0h^{-1}$Mpc for a fixed value
of $\gamma=1.6$, equivalent to the value reported in A03 and
\citet{a05} for the LBG auto-correlation.  Our decision to center the
DLAs in the observed fields prevented an estimation of the DLA-LBG 
cross-correlation effectively beyond $\sim4h^{-1}$Mpc using the above
method.  However, our cross-correlation values are consistent with the 
constraints placed on the DLA-LBG cross-correlation by \citet{b04}
using a comparable analysis over a range of $\sim5-15h^{-1}$Mpc.

\subsection{Maximum likelihood}

\begin{figure}
\begin{center}
%\scalebox{0.6}[0.6]{\rotatebox{90}{\includegraphics{f2.ps}}}
\scalebox{0.6}[0.6]{\rotatebox{90}{\includegraphics{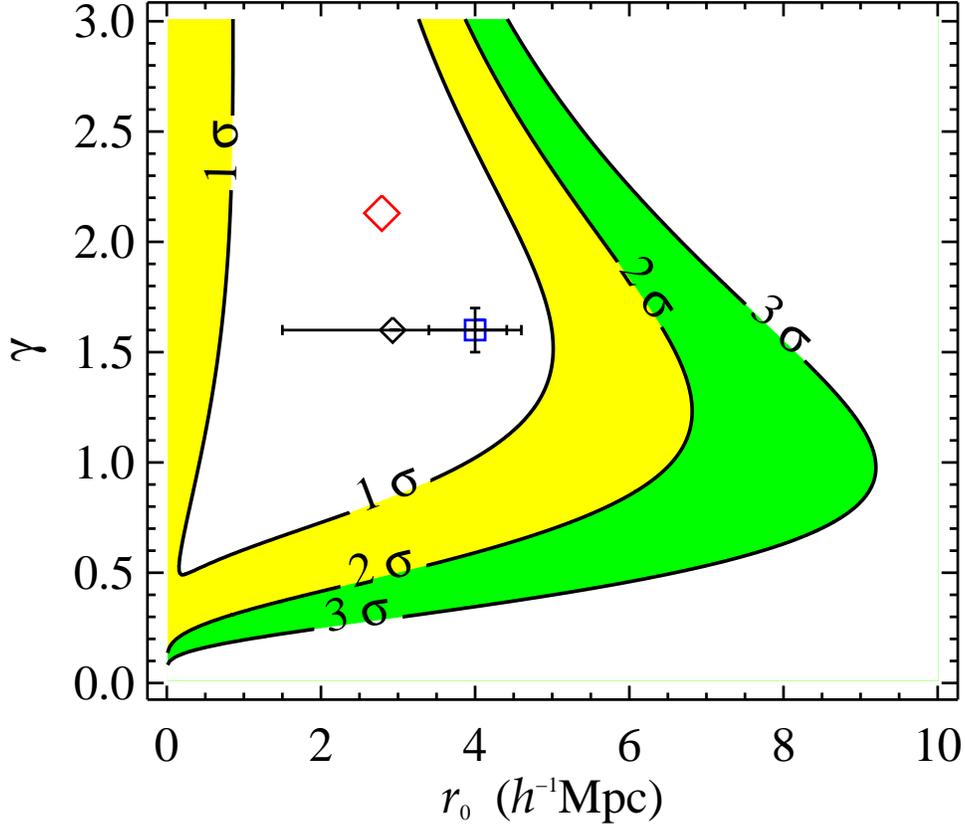}}}
\caption
{Two parameter probability contours for the DLA-LBG cross-correlation
  using the maximum likelihood method.  The best fit values of
  $r_0=2.8^{+1.4}_{-2.0}h^{-1}$Mpc$,~\gamma=2.1^{+1.3}_{-1.4}$ are
  indicated by the large red diamond.  The best fit value of
  $r_0=2.9^{+1.4}_{-1.5}h^{-1}$Mpc for a fixed value of $\gamma=1.6$
  (small black diamond) is shown with the associated $1\sigma$
  uncertainty on $r_0$. For comparison, the blue square and error bars
  indicate the LBG auto-correlation best fit values and $1\sigma$
  uncertainties of $4.0\pm0.6h^{-1}$Mpc$,~\gamma=1.6\pm0.1$ from
  Adelberger et al. (2005).  Here, the angular positions of the
  galaxies in the random catalogs were made to be identical to the data
  to minimize possible artificial clustering effects caused by the
  physical constraints of the slitmasks.}
\label{fig:ml}
\end{center}
\end{figure}

As an independent method of analysis, and to make the most of our
dataset, we determined the maximum likelihood of a power law fit 
(equation~\ref{powerlaw}) to the observed data
\citep{croft97,mullis04}.  We divided the radial separations into a
large number of finely spaced regular intervals that coincided with
either 1 or 0 LBGs.  Poisson statistics hold in the regime of large 
interval number and small probability per interval.  We used this 
to form the likelihood function
\begin{equation}
{\cal L} = \prod_i^N \frac{e^{-\mu_i} \mu_i^{\nu_i}}{\nu_i !} 
\prod_{j\ne i}^N  \frac{e^{-\mu_j} \mu_j^{\nu_j}}{\nu_j !}
\end{equation}
where $\mu_i$ is the expected number of pairs in the $i$th interval,
$\nu_i$ is the observed number of pairs for that same interval, and
the index $j$ runs over the elements where there are no pairs.  
The expected number of pairs was determined by solving
equation~\ref{DLAest} for $D_{DLA}D_{LBG}$ over a reasonable range of
$r_0$ and $\gamma$.  We then maximized the expression $S = -2 \ln
{\cal L}$.  Confidence levels were defined as $\Delta S =
S(r_{0,best},\gamma_{best}) - S(r_0, \gamma)$ with the assumption that
$S$ has a $\chi^2$ distribution.  We found the best fit values and
68\% confidence levels for the cross-correlation using this method to
be $r_0=2.8^{+1.4}_{-2.0}h^{-1}$Mpc,$~\gamma=2.1^{+1.3}_{-1.4}$ and 
a best fit value of $r_0=2.9^{+1.4}_{-1.5}h^{-1}$Mpc for a fixed
$\gamma=1.6$.  Figure~\ref{fig:ml} displays these results.

\citet{c05b} describe the above analyses in more detail, present
several tests to address the shortcomings of each method, and make
efforts to quantify the physical effects that the multi-object
slitmasks have on the clustering signal.  A short summary of best fit 
values and $1\sigma$ uncertainties described here and from that work
is listed in Table~\ref{table:DLresults}.  It can be seen that all 
independent methods, and tests thereof, result in consistent central
values within their uncertainties.  

\section{Discussion}

The LBG bias at $z\sim3$, derived from the LBG auto-correlation
of the R $<25.5$ spectroscopic sample, indicates an average
LBG dark matter halo mass of $\sim10^{12}M_\odot$ \citep{s98,a98}.
The 11 DLAs presented here comprise an unbiased representation of a
random cross-section weighted sample.  The similarity between the LBG
auto-correlation and the DLA-LBG cross-correlation implies that the
bias factors of the two populations are comparable and that DLAs, on 
average, may form in similarly massive potential wells.  We consider
the implications from the best fit central value $r_0\sim2.9$ using
the maximum likelihood method with fixed $\gamma=1.6$.  This
measurement corresponds to a DLA galaxy bias of $b_{DLA}\sim2.4$ and 
represents an average DLA halo mass of $\langle
M_{DLA}\rangle\sim10^{11.2}M_\odot$.  This value is higher than what
is predicted for DLAs by simple models \citep{m98} but is in very good 
agreement with numerical models of varying resolution that invoke
strong galactic-scale winds \citep{n05} and thermal feedback 
\citep{maller01,b05}.  In these models, galactic outflows purge
low-mass halos of gas capable of generating damped Lyman $\alpha$ 
absorption lines and result in a higher mean mass.  It must be noted
that because CDM predicts a steeply rising mass function, it is
expected that the median DLA halo mass is lower than the mean mass
implied here.  

The observed similarity in the spatial distributions of DLAs and LBGs
also helps support previous ideas that the two populations may be
connected \citep{schaye01,m02}.  Recent analysis of the {\it Hubble
Ultra Deep Field} \citep{cw05} shows little evidence for {\it in situ}
star formation throughout the DLA neutral gas.  However, calculations 
using the \textsc{Cii*} absorption feature to trace the cooling rates
in DLAs \citep{w03} require local sources of radiation to heat DLAs.
A plausible scenario is that LBGs are embedded in the same systems
that contain DLAs.  This model is reinforced by the near equality
between DLA cooling rates and LBG heating rates.  Furthermore, the
lack of detected DLA emission out to reasonable impact parameters from 
background QSOs and the above implied average halo mass are consistent
with DLAs sampling the bulk of the LBG population with typical impact 
parameters of $<1\arcsec$ and median luminosities of R $>27$. 

Although the uncertainties in this initial measure of the
three-dimensional DLA-LBG cross-correlation function are large, there 
remains strong evidence for an overdensity of LBGs near DLAs.  The
similarity in the central values between the DLA-LBG cross-correlation
and the LBG auto-correlation underscore a need for additional
observations to improve the statistical significance of these results
and to provide a step toward a complete picture of galaxy formation
and evolution.

\acknowledgments
We would like to thank K. L. Adelberger and C. Mullis for helpful 
and informative discussions.  This work was partially supported by the
National Science Foundation grant AST-0307824 and the NSF Astronomy \& 
Astrophysics Postdoctoral Fellowship (AAPF) grant AST-0201667 awarded
to Eric Gawiser.

%---------------------------------------------------------------------------
%   FIGURES
%--------------------------------------------------------------------------- 
\clearpage

\clearpage

\clearpage

%---------------------------------------------------------------------------
%   TABLES
%---------------------------------------------------------------------------
\begin{deluxetable}{lcc}
\tablecaption{DLA-LBG Cross-Correlation Parameter Summary 
\label{table:DLresults}}
\tablewidth{0pt}
\tablehead{
\colhead{Method} & \colhead{$r_0$} & \colhead{$\gamma$}}
\startdata
Conventional binning\tablenotemark{a}\tablenotemark{b}
                        & $3.32\pm1.3$ & $1.74\pm0.4$ \\
Maximum likelihood\tablenotemark{b}\tablenotemark{c}
                        & 2.81$^{+1.4}_{-2.0}$ & 2.11$^{+1.3}_{-1.4}$ \\
Cumulative $\chi^2$ test\tablenotemark{b}\tablenotemark{c}\tablenotemark{d}
                        & 3.84$^{+4.2}_{-3.8}$ & 2.06$^{+2.0}_{-1.3}$ \\
\hline
Conventional binning\tablenotemark{a}\tablenotemark{d}\tablenotemark{e}    
                        & $3.21\pm1.0$ & $2.03\pm0.2$ \\
Maximum likelihood\tablenotemark{c}\tablenotemark{d}\tablenotemark{e}   
                        & 3.20$^{+2.2}_{-2.9}$ & 1.62$^{+1.4}_{-1.0}$ \\
Cumulative $\chi^2$ test \tablenotemark{c}\tablenotemark{d}\tablenotemark{e}
                        & 3.91$^{+4.4}_{-3.9}$ & 2.11$^{+2.7}_{-1.3}$ 
\enddata
\tablenotetext{a}{Galaxy separations binned using the cylindrical
  approach described in Adelberger et al. (2003), Appendix C}
\tablenotetext{b}{Angular positions of galaxies in the random
  catalogs are identical to the angular positions of the data (to
  minimize possible artificial clustering effects caused by the
  slitmasks)}
\tablenotetext{c}{Galaxy separations binned radially}
\tablenotetext{d}{Described in \citet{c05b}}
\tablenotetext{e}{Angular positions of galaxies in the random 
  catalogs are random}
\end{deluxetable}

\end{document}